\documentclass[english]{article}
\usepackage{geometry}
\usepackage[T1]{fontenc}
\usepackage[latin1]{inputenc}
\usepackage{graphicx}
\makeatletter
\usepackage{babel}
\makeatother
\begin{document}
\title{Anomalous frequency shifts in the solar system}
\author{Jacques Moret-Bailly}

\maketitle
jacques.moret-bailly@u-bourgogne.fr

\begin{abstract}
The improvements of the observations of the solar system allowed by the use of probes and big instruments let appear several problems: The frequencies of the radio signals received from the probes sent over  5 UA from the Sun are too high; the explanation by spicules or  siphon-flows of the frequency shifts of UV emissions observed on the surface of the sun by SOHO is not satisfactory; the anisotropy of the CMB seems bound to the ecliptic.

This problems are solved using a coherent optical effect, deduced from standard spectroscopy and easily observed with lasers. In a gas containing atomic hydrogen in states 2S and (or) 2P , transfers of energy between light beams, allowed by thermodynamics,  produce the required frequency shifts or amplifications.
\end{abstract}

\section{Introduction}
The remarkable precision of relastivistic mechanics and electrodynamics allows, for instance the good localizations by the the Global Positioning System. However, a discrepancy appears in the observation of the probes (in particular Pioneers 10 and 11) when their distance from the Sun becomes larger than about five astronomical units. The the frequencies of the received radio signals are too high, so that it seems that the attraction by the Sun increases over Newton's law. Several prudent explanations are proposed, in particular new physics or an acceleration by an anisotropic radiation of the energy provided by the disintegration of the plutonium which feeds the probes in energy.

In section \ref{description}, we show why the previous explanations cannot work, showing that the problem occurs during the propagation of the radio waves.

Other discrepancies are found in observations of the Solar system:

 - the explanations of the redshifts of the UV emission spectra of the Sun observed by SOHO, by spicules or siphon-flows, appear weak; 

-  the anisotropy of the microwave background appears bound to the Solar system.

This section defines the properties required for an optical effect able to solve these three problems.

In section \ref{creil}, we study this effect, deduced from standard rules of spectroscopy, but observable only in conditions  which allow to qualify the light pulses  ''ultrashort''. This effect appears, in particular while light is refracted by a low pressure gas containing atomic hydrogen in states 2S or 2P.

\section{Description of the anomalies and their explanations.}\label{description}
\subsection{Anomalies of the speeds of the Pioneer 10 and 11 probes.}\label{Pioneer}
The original description of the Pioneer probes, and of the detection of anomalies in the radio-signals of several probes was given by Anderson et al. \cite{Anderson1,Anderson2}.

Anderson et al. deduce the radial speed of the probes through an assumed Doppler shift of radio waves: An electromagnetic wave is sent from the Earth to the probes at a frequency deduced from the frequency of an hydrogen maser by a multiplication such that the frequency received by the probe is close to 2.11 GHz. This frequency is multiplied by 240/221 to avoid an interference with the received frequency, amplified and sent back to the Earth where it is detected by an heterodyne system, producing a frequency close to 1 MHz. The weakness of the received signal requires a track more and more difficult with an increase of the distance.

\begin{figure} \includegraphics[height=10cm] {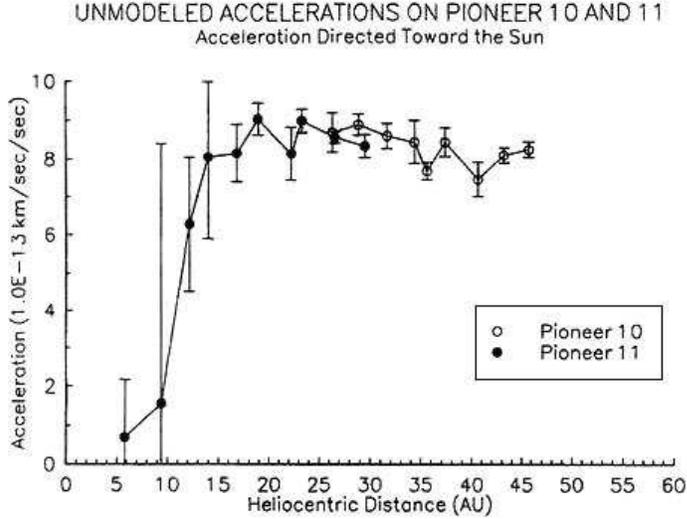}
 \caption{ Apparent acceleration corresponding to the residual frequency shift, as a function of distance of the Sun, from Anderson et al. \cite{Anderson2}.}
\label{F1}
\end{figure}

Taking into account the main computed frequency shifts, Doppler and gravitational, less important perturbations such as the pressures of radiation of the Solar light, and the pressure of the Solar wind, the gravity of the Kuiper belt ..., the received frequency has the computed value until the distance of the probe is lower than 5AU; at a longer distance the received frequency becomes more and more too high, until the extra acceleration stabilises  over 15 AU at the value $(8.6  \pm 1.34)\times 10^{-8} $ cm s$^{-2}$. See figure \ref{F1}.

If the origin of the acceleration were a leakage of the valves of the thrusters  allowing the maneuvers, the probability that both Pioneers have leaks producing the same acceleration, and that a leak reproduces after a maneuver, is low. Therefore, the main hypothesis is an anisotropy of the radiation of the 2 kW produced by the decay of the plutonium on board the aircraft. The decrease of this energy with the time is not observed, but it may correspond to the uncertainty of the measure of the acceleration\cite{Markwardt,Scheffer}.

We think that the origin of the anomalous accelerations does not lie in the apparatus for the following reasons :

i)The identities of the accelerations of both Pioneers show that they do not probably result from an accidental disworking such as a leakage of a valve.

\begin{figure} \includegraphics[height=16cm,angle=270] {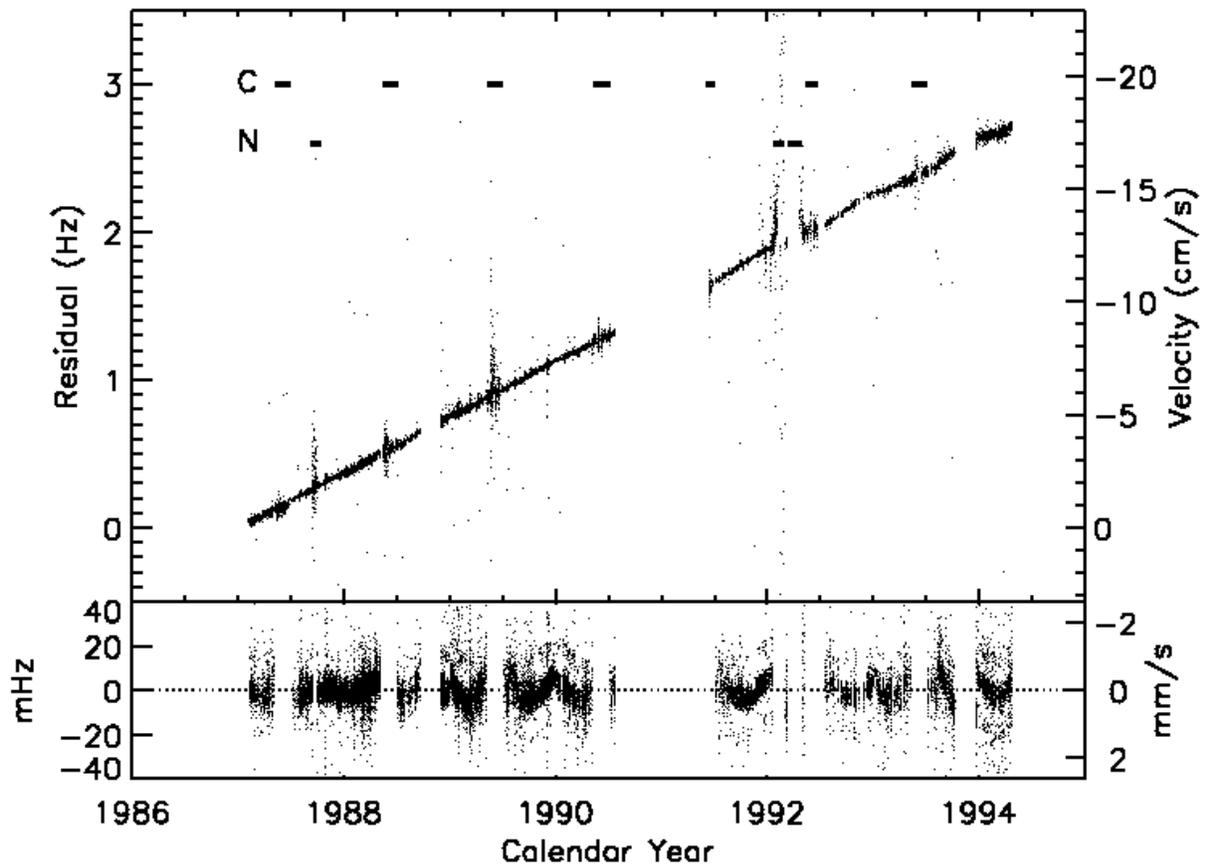}
 \caption{ Doppler residuals as  a function of time.  The top panel shows all of the data. The bottom shows the residuals, excluding  the regions perturbed by the solar corona, designated by an horizontal bar "C"  and the noisy regions,  designated "N".  From Markwardt\cite{Markwardt}.}
\label{F2}
\end{figure}

ii)On figure \ref{F2}, the interferences with the corona produce large perturbations of the observed frequencies ("C" regions), but , after, the linear increase of speed is restored.. If the "N" regions were produced in the apparatus, the large anomalous speeds which would appear should translate the following segments; thus something similar to a path through the corona, happens on the path of the light, and the properties of this path are more easily restored than the properties of a complex apparatus.

\medskip
What can happen on the path of light in the Solar wind ? Maybe the proximity of a planet, maybe an increase of the Solar activity. How do this change in the Solar wind may be transferred to the radio signal ?

\subsection{The redshifts of the UV emission lines of the quiet Sun.}\label{sun}
The chromosphere of the  quiet Sun was studied by Peter and Judge \cite{Peter}  using data acquired by the Solar Ultraviolet Measurement of Emitted Radiation (SUMER), on the SOHO spacecraft. 

We consider here only residual frequency shifts obtained by subtraction from the observed shifts of a "main correction'': a) the the Doppler shift produced by the rotation of the Sun and the relative movement of the Sun and the probe; b) the relativistic shift. 

\medskip
After a description of spectra, Peter \& Judge present the current state of their interpretation, founded on an attribution of the (residual) frequency shifts of the spectral lines to a Doppler effect produced by vertical movements of the gas in the chromosphere. 

To explain that lines emitted at the same, or at very close places have different redshifts, an hypothesis is that gas is ejected in vertical spicules, then cools and flows down; an other hypothesis is siphon flows through loops. But Peter \& Judge write : ''As for the spicule idea, the existing siphon-flow pictures are either non valid or only part of the story''. Other hypothesis are tried, but "still more work is needed''.

With the hypothesis of Doppler effects and vertical movements, for all lines, there is no (residual) frequency shift at the limb of the Sun. This hypothesis implies that the frequencies measured at the limb are, after subtraction of the "main correction'', the absolute frequencies. Comparing the absolute frequencies deduced from SUMER measures at the limb to older measured or computed frequencies, discrepancies appear, attributed to a lack of precision of the old results. For instance,  a computed value of the wavelength of Mg X is 62495.2 pm, while the value deduced from the observation of the limb is 62496.8 $\pm$ .7 pm.

The wavelength of the  Ne VIII line was measured in the laboratory by Bockasten et al \cite{Bockasten} who found 77040.9 $\pm$ .5 pm. From  SUMER measures, at the limb, Peter \& Judge obtained 77042.8 $\pm$ .7 pm. Considering that this value is a rest wavelength, there is a discrepancy attributed to  a too short error bar in the laboratory measure. Peter \& Judge write : "If one would take the Bockasten et al. value for granted, this would imply that the Ne VIII is indeed redshifted at disk center and would beg the question of how the redshift of a line seen at disk center $C$ can even increase toward the limb - - we would not be able to explain such a variation with our current understanding of the solar atmosphere.'' 

Peter \& Judge  do not rely much on the theory they use: ''Neither the nature of the driving motions nor the response of the plasma can be reliably constrained by currently available observations or by numerical simulations ... It might be that the blueshifts we observe are not caused by the out-flowing solar wind but by some other processes.''

An other process, a new understanding is supposing that the shifts occurs during the propagation of the light through a shell of the chromosphere: the path through this shell for the rays emitted at $C$ is the thickness of the shell, while it is larger for other rays. Writing $S(M)$ the frequency shift obtained by Peter \& Judge at a point $M$, $s(M)$ the newly defined shift, $L$ being a point at the limb, the relation between the shifts is:

\begin{equation}
S(M) = s(M)-s(L) \label{frsh}
\end{equation}

As $|s(L)|$ is larger than $|s(M)|$, the signs of  $S(M)$ and $s(M)$ are opposite, the variations of the frequency shifts along a radius are opposite.

\begin{figure} \includegraphics[height=12cm] {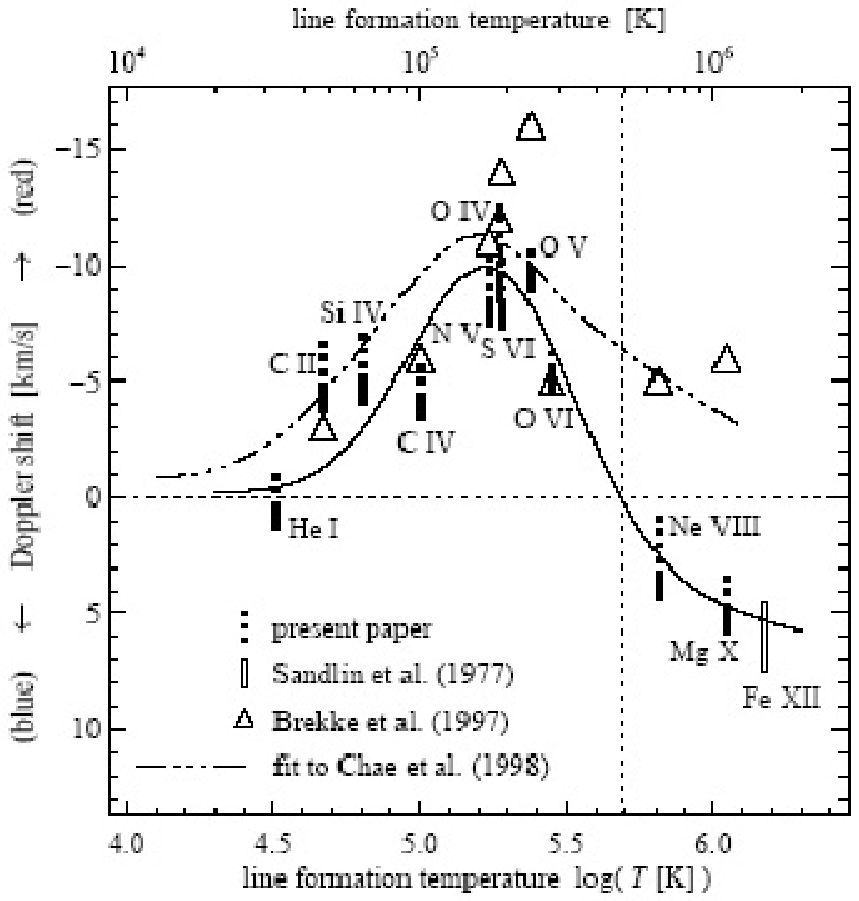}
 \caption{Variation of the frequency shift $S(C)$} with formation temperature of  the line. Error bars for the data of Brekke et al (1997) were typically 2 km s$^{-1}$. The solid line is a by-eye fit of  the Doppler Shifts in Peter \& Judge study. From Peter \& Judge \cite{Peter}.
\label{pj} 
\end{figure}

Figure \ref{pj} shows  the shifts of various lines $S(C)$ as a function of the temperature of the emitting gas. Suppose that the column density is sufficient to reach nearly a saturation, that is an equilibrium between the temperature of the emitting gas and the temperature of the light \footnote{Temperature deduced from the intensity in a mode, using Planck's formula for the radiation of a black body.} at the centre of the lines. Thermodynamics says that energy flows from hot to cold, so that the three high energy lines Ne VIII, Mg X and Fe XII are allowed by thermodynamics to transfer energy to the other lines provided that the light is refracted by a convenient medium playing the role of a catalyst. This transfer  redshifts the three hot lines, and blueshifts the other in conformity with the definition $s(M)$ of the redshifts.

Why do the lines emitted at the lowest temperatures (He I, C II, Si IV and C IV) are less blueshifted than the other cold lines? A simple explanation is that the catalytic power of the refracting medium is nearly zero at temperatures lower than 30 000 K (temperature of emission of He I) and gets a maximal mean value if the temperature along the path varies from about 170 000 K (maximum of the curve on figure \ref{pj}) to 30 000 K,  leading to an optimal temperature of the effect, very roughly, of the order of  100 000 K.

\subsection{The anisotropy of the cosmic microwave background.}

In subsection \ref{Pioneer}, we explained the anomalous increase of frequency of the Pioneer probes by an interaction in the solar wind. If this interaction is similar to the interaction whose characteristics were found in subsection \ref{sun},  it is a transfer of energy from the solar light to radio waves. This transfer applies to all radio waves propagating in the solar wind over 5 UA, in particular to the cosmic microwave background.

The solar  wind is generated in the holes of the corona, so that it is anisotropic. Its structure may be modified by the magnetic fields of the planets. Thus, the blueshift of the radio frequencies by the solar wind is anisotropic. For the CMB, a thermal radiation, this shift is an amplification which adds a contribution to the anisotropy  due to the movement of the Sun in the galaxy.  The analysis of the observed CMB leads to a similar result \cite{Schwarz,Land,Naselsky}.

\section{The Coherent Raman Effects on Incoherent Light (CREIL)}\label{creil}
A simultaneous explanation of the anomalies studied in section \ref{description} uses an optical effect having the following properties:

i)The images and the spectra are not blurred; else the signals from the Pioneers would be too much weakened;

ii)The energy transferred from hot beams to cold beams shifts the frequencies;

iii)The interacting beams must be refracted by a gas whose optimal temperature is of the order of 100 000 K; observed from the solar frequency shifts and the cooling of the solar wind.

\medskip

This section explains the required effect, which appears very similar to the refraction, but which requires very particular media, or ultra-short laser pulses.

\subsection{Conditions for Doppler-like frequency shifts by interaction with matter.}\label{con}

 - A Doppler-like redshift must avoid a blur of the images. Therefore, it must be space-coherent, so that the wave surfaces are not disturbed:
For an involved molecule, it exists relations between the local phases of all involved electromagnetic fields, and the phases of all molecular oscillators;
\textquotedblleft space coherence'' means that these relations are identical for all involved molecules. Consequently, supposing that the number of involved molecules is large, Huygens' construction shows that the radiated fields generate clean wave surfaces related with the wave surfaces of the exciting fields.

 - For a time-coherent source (continuous wave laser), \textquotedblleft frequency shift'' means that while the source emits $n$ cycles, the detector receives a different number $m$. Thus, the number of cycles between the source and the receiver is increased of $n-m$; it is an increase of the number of wavelengths, thus an increase of the distance, therefore a Doppler effect. Consequently, a Doppler-like redshift is only possible with time-incoherent light; a parameter measuring this incoherence must appear in the theory to forbid an application to time-coherent light.

 - The energy absorbed by the redshifting process  must not be quantised to avoid a blur of the spectra:
If a light beam exchanges a quantified energy with a molecule, a fraction of the intensity of the beam gets a finite shift. In a parametric process, the molecules leave their stationary state only temporarily, their states becoming \textquotedblleft dressed'' during their interactions with the light; the light beams exchange not-quantified energy, the matter plays the role of a catalyst\footnote{We do not follow an extended definition of \textquotedblleft parametric''  interactions in which the matter may be (des)excited during the interaction (for instance in a He-Ne laser medium), \textquotedblleft parametric'' becoming synonymous of \textquotedblleft coherent''.}.

\subsection{Reminding the semi-classical theory of refraction.}

\medskip
{ \bf Macroscopic theory.}

To simplify the explanations, suppose that the refracting medium is perfectly transparent.

 A sheet of matter between two close wave surfaces distant of $\epsilon$ is excited at a pulsation $\Omega$. The sheet radiates a Rayleigh coherent wave late of $\pi/2$ whose amplitude is a small fraction $K\epsilon$ of the exciting amplitude $E_0$. From Huygens' construction it generates the same wave surfaces, so that the fields add into 

\begin{eqnarray}
E=E_0[\sin(\Omega t)+K\epsilon \cos(\Omega t)]\nonumber\\
\approx E_0[\sin(\Omega t)\cos(K\epsilon)+\sin(K\epsilon )\cos(\Omega t)]=E_0\sin(\Omega t -K\epsilon).
\end{eqnarray}
 This result defines the index of refraction $n$ by the identification 

\begin{equation}
K=2\pi n/\lambda=\Omega n/c.\label{refr}
\end{equation}

\medskip
{\bf Microscopic, quantum theory.}

Suppose that the light interacts with free identical molecules, initially in the same non-degenerate stationary state $\phi_0$. The perturbation of a molecule by an electromagnetic wave mixes $\phi_0$ with other states $\phi_i$ , producing a non-stationary state $\Phi = C_0\phi_0+\sum_iC_i\phi_i$, where the $C_i$ are very small.

We must consider the set of all interacting molecules, adding an upper index $k$ to distinguish the molecules. Without a field, the total, stationary state is $\Psi_0=\prod_k\phi_0^k$. Its degeneracy is the number of molecules.

Perturbed by an external field,  the refracting medium radiates a scattered, coherent field late of $\pi/2$, generating the same wave surfaces than the exciting field; therefore,  the dynamically excited, non-stationary, \textquotedblleft dressed '' ( or  \textquotedblleft polarisation'' ) state $\psi^m$ which emits this field is  characterised by an index $m$ representing the exciting mode. 

Considering other refracted modes, $\Psi$ splits as $\prod_m\psi^m$.

\medskip

Remark that the coherent interactions are much stronger than the incoherent: A refraction by $\approx 0.25 \mu m$ of water delays the light of $\pi/2$, that is the light is fully scattered by the coherent Rayleigh scattering. In a swimming pool, we see well through 25 metres of water, only a fraction of the light is scattered by the incoherent Rayleigh scattering; the factor is $10^8$. 

\subsection{ Principle of the CREIL.}

The CREIL results from an interaction between dressed states $\psi^m$; as these states have the same parity, the interaction must be of Raman type, for instance quadrupolar electric. Thermodynamics says that the entropy must increase, so that the floods of energy are from the modes which have a high Planck's temperature  to the colder ones. For an astrophysical application we consider a purely parametric effect: the matter, a low pressure gas in low fields, returns to its initial state after an interaction. 
 
The dressed state $\psi^m$ radiates a mixture of the coherent Rayleigh scattering which produces the refraction and coherent Raman scatterings. These locally weak scatterings may be studied independently, so that the CREIL may be considered as a set of {\it simultaneous} Stokes and anti-Stokes coherent Raman scatterings with a zero balance of energy for the molecules \footnote{\textquotedblleft Coherent Raman Scatterings on Incoherent Light'' (CREIL) is ambiguous, relative either to a single Raman interaction ( ignoring the quasi-resonant, easy transfer of the Raman energy to the thermal radiation ), or to the whole set of interactions.}. The scattered beams have the same wave surfaces than the exciting beams, so that these beams may interfere, as in the coherent Rayleigh scattering making the refraction; as the scattered fields are much weaker than the exciting field, they may be added independently to it. The pulsations of the Raman beams are shifted by $\pm\omega$, and, at the beginning of a pulse, in phase because the resonance introduces a $-\pi/2$ phaseshift. The sum of the exciting wave and the coherent anti-Stokes scattered wave is:

\begin{eqnarray}
E=E_0[\sin(\Omega t)+K'\epsilon \sin((\Omega+\omega)t)] \hskip 3mm ({\rm with } (K'>0) 
\nonumber\\ 
E=E_0[\sin(\Omega t)+K'\epsilon[\sin(\Omega t)\cos(\omega t)+\sin(\omega t)\cos(\Omega t)]].
\end{eqnarray}
Supposing that $\omega t$ and $K'\epsilon$ are small, the second term, product of two small quantities, may be neglected, and the last one transformed:

\begin{eqnarray}
E\approx E_0[\sin\Omega t+\sin(K'\epsilon\omega t)\cos(\Omega t)]\nonumber\\
E\approx E_0[\sin(\Omega t)\cos(K'\epsilon\omega t)+ \sin(K'\epsilon\omega t)\cos(\Omega t)=E_0\sin[(\Omega+K'\epsilon\omega)t].\label{eq4}
\end{eqnarray}
$K'\epsilon$ is an infinitesimal term, but the hypothesis $\omega t$ small requires that the Raman period $2\pi/\omega$ is large in comparison with the duration of the experiment $t$.

This condition was set by G. L. Lamb Jr. for the definition of \textquotedblleft ultrashort pulses'' : \textquotedblleft shorter than all relevant time constants'' \cite{Lamb}. With ordinary light, the time coherence plays the role of length of the pulses:  thus, the time-coherence, some nanoseconds, must be \textquotedblleft shorter than all relevant time constants''.

We have found a first relevant time constant. A second is the collisional time constant, because the collisions destroy the space-coherence, producing an ordinary, weak, incoherent Raman scattering; a low pressure gas is needed. 

The same computation, replacing $K'$ by a negative $K''$ gives the Stokes contribution, so that we replace $K'$ by $K'+K''$ in formula \ref{eq4}. $K'+K''$ depends on the difference of population in both levels, that is on $\exp(-h\omega/2\pi kT)-1\propto \omega/T$, where $T$ is the temperature of the gas.

The theory of the refraction shows that the index of refraction is nearly constant in the absence of resonance close to $\Omega$, so that, using for the polarisability a formula equivalent to formula \ref{refr}, { } $K'+K"$ { }appears nearly proportional to $\Omega \omega/T$, and the frequency shift is :

\begin{equation}
\Delta\Omega=(K'+K")\epsilon\omega\propto \epsilon\Omega\omega^2/T.
\end{equation}

 The relative frequency shift $\Delta\Omega/\Omega$ is nearly independent on $\Omega$.

All required properties are obtained: space coherence, limitation of the time-coherence, no excitation of the gas, nearly constant relative  frequency shift. As the shift is proportional to $\omega^2$, a strong effect requires a Raman pulsation $\omega$ as large as allowed by the preservation of the coherence. As the time-coherence of ordinary light is some nanoseconds, an  Raman frequency is of the order of 100 Mhz.

\subsection{Laboratory observation of the CREIL effect}
Usually, it is not necessary to take into account the radiations which receive energy because we are surrounded by thermal radiations whose blueshift is simply a heating. In a convenient medium, the CREIL effect transfers also energy between the radio frequencies which make the thermal radiation as long as the thermal equilibrium, including the isotropy, is not reached; this CREIL effect is strong because, all involved frequencies being low, it is nearly resonant, so that the radio frequencies get quickly a thermal equilibrium.

The CREIL in optical fibres is so easily obtained that it makes problems for the use of short pulses in telecommunications. With the high peak power of femtosecond lasers, the index of refraction and the components of the tensor of polarisability become increasing functions of the intensity, allowing a study of the effect in small cells. This nonlinear effect named "Impulsive Stimulated Raman Scattering" (ISRS) allows an easy study of the properties of the coherent Raman effect on incoherent light: transfer of energy from a laser beam to another producing frequency shifts, verification of Lamb's conditions (Yan et al. \cite{Yan}).

While the lengths of the laser pulses increase, the experiments become more and more difficult: To increase the collisional time, it becomes impossible to use dense matter, a gas less and less dense must be used. While it is easy to find strong Raman resonances at the rotational and vibrational frequencies of molecules, resonances close to 100 MHz appear generally in highly excited states, almost unpopulated. Therefore, an observation of a CREIL effect, using ordinary incoherent light would require an expansive experiment while it is well verified in the whole easily accessible domain of frequencies. 

\subsection{Propagation of incoherent light in atomic hydrogen}\label{H}

As atomic hydrogen has a simple spectrum, its levels of energy may be well populated. Its electric quadrupole spin recoupling transition ($\Delta F=1$) in the ground state has the frequency 1420 MHz, too high. But, in the first excited state, the frequencies 178 MHz in the 2S$_{1/2}$ state, 59 MHz in 2P$_{1/2}$ state, and 24 MHz in 2P$_{3/2}$ are very convenient; in these states, the gas will be named H*. It is more difficult to populate higher states, and the resonance frequencies are low, so that, in these states, the CREIL effect is negligible.

\medskip
Excited atomic hydrogen which redshifts the light may be generated by various processes:

\medskip
{\bf Thermal excitation of hydrogen.}

The ionisation energy equals $kT$ for a temperature $T=156 000 K$; as the energy needed for a pumping to the  states of principal quantum number $n=2$ (H* states) is the three fourth of the ionisation energy, it equals $kT$ for $T=117 000 K$. Using Boltzman law, these temperatures may be considered as indicating roughly where these particular states of hydrogen are abundant, remarking however that by a thermal excitation, the proportion of hydrogen in the H* states is  limited by the excitation to higher values of $n$, and by the ionisation at low pressures. Remark that, from figure \ref{pj}, we found in \ref{sun} an approximate optimal value $T=100 000 K$ : H* is clearly the source of the anomalous frequency shifts on the Sun.

\medskip
{\bf Lyman $\alpha$ pumping of atomic hydrogen.}

Over a temperature $T=10 000 K$, the molecules of hydrogen are dissociated. The strong absorption of the Lyman alpha line produces H*.  The effective decay of H* is very slow at low pressures because this decay can only re-emit the Ly$_\alpha$ line which is strongly, immediately re-absorbed. The surface of the Sun is too cold to provide much energy  at the Ly$_\alpha$ frequency. But H* may be produced close to very hot objects such as quasars, accreting neutron stars. A feed-back may appear in unexcited atomic hydrogen illuminated by a far UV continuous spectrum: The excitation at the Lyman $\alpha$ frequency produces H*, therefore a redshift which renews the intensity of the light at the Lyman $\alpha$ frequency until a previously absorbed line almost stops the redshift, so that the other Lyman lines are strongly absorbed and will nearly stop the following fast redshift. 

\medskip
{\bf Cooling of an hydrogen plasma.}

The combination of the protons and electrons of a plasma produces atomic hydrogen in various states of excitation. The 2S state is stable at a low pressure. The optical transitions from the 2P states generate a Ly$_\alpha$ line which may be reabsorbed. The cooling of the solar wind beyond 5 UA produces H* and explains the blueshift of the radio-frequencies of the Pioneers 10 and 11,  at least a part of the anisotropy of the CMB bound to the ecliptic.

\section{Conclusion}
Introducing coherent optical interactions other than the refraction seems the key of a lot of explanations of up to now difficult to understand astrophysical observations. In particular, the Coherent Raman Effects on Incoherent Light (CREIL)  is the true origin of frequency shifts usually considered as produced by a Doppler effect. The use of the CREIL is very simple:  light beams refracted simultaneously by a gas containing atomic hydrogen in states 2S or 2P exchange energy to increase the entropy of their set, producing frequency shifts. Where the physical conditions allow the production of H*, anomalous frequency shifts appear.

\end{document}